\documentclass{article}
\usepackage{amssymb, amsfonts, amsmath}
\usepackage{graphicx}
\usepackage{enumerate, setspace}
\usepackage{natbib}
\usepackage{url} 
\usepackage{multirow}
\usepackage{xcolor}

\newcommand{\argmin}{\operatornamewithlimits{argmin}}

\title{CompModels: A suite of computer model test functions for Bayesian optimization}
\author{Tony Pourmohamad\thanks{
    Corersponding author. E-mail: tpourmohamad@gmail.com} \\ \emph{Genentech, Inc.}}

\begin{document}

\maketitle 

\begin{abstract}
 The{ \bf CompModels} package for \textsf{R} provides a suite of computer model test functions that can be used for computer model prediction/emulation, uncertainty quantification, and calibration, but in particular, the sequential optimization of computer models. The package is a mix of real-world physics problems, known mathematical functions, and black-box functions that have been converted into computer models with the goal of Bayesian (i.e., sequential) optimization in mind. Likewise, the package contains computer models that represent either the constrained or unconstrained optimization case, each with varying levels of difficulty. In this paper, we illustrate the use of the package with both real-world examples and black-box functions by solving constrained optimization problems via Bayesian optimization. Ultimately, the package is shown to provide users with a source of computer model test functions that are reproducible, shareable, and that can be used for benchmarking of novel optimization methods. 
\end{abstract}


\section{Introduction}\label{sec:intro}
The {\bf{CompModels}} package \citep{CompModels} for \textsf{R} \citep{R} is a suite of test functions designed to mimic computer models. Usually deployed when physical experimentation is not possible, a computer model (or code) is a mathematical model that simulates a complex phenomena, or system, under study via a computer program. For example, weather phenomena, such as hurricanes or global warming, are not reproducible physical experiments, therefore,  computer  models  based  on  climatology  are  used  to  study  these  events. At its simplest, a computer model is a mathematical model of the form
\begin{align}
y = f(x_1,\dots,x_d) = f(x), \,\,\,\,x=(x_1,\dots,x_d)^T\in\mathcal{X},
\end{align}
where $x$ is an input variable to the computer model, $y$ is a (possibly multivariate) deterministic output from the computer model, and $\mathcal{X}$ is the domain of the input variable. A defining characteristic of most computer models is that, for a given input $x$,  the evaluation of the underlying mathematical model, $f$, is a time intensive endeavor. Computationally expensive computer models helped spur the development of the computer modeling field in statistics \citep{sant:will:notz:2003}, and in particular, the development of ``cheap-to-compute" statistical models,  or \emph{surrogate  models},  that  resemble  the  true  computer model very closely but are much faster to run. Outside the scope of this paper, but useful for forthcoming discussion and illustrations, we simply mention that Gaussian processes (GPs) \citep{stein:1999} have been used as the typical modeling choice for building statistical surrogate models, and this is due to their flexibility, well-calibrated uncertainty, and analytic properties \citep{gramacy:2020}.

Another typical trait of computer models is that they are often treated as black-box functions. Here, a black-box computer model is a computer model where evaluation requires running computer code that reveals little information about the functional form of the underlying mathematical function, $f$. The black-box assumption often arises due to the fact that $f$ may be extremely complex, analytically intractable, or that access to the internal workings of the computer model are restricted, say, for such reasons as being proprietary software. The latter restricted cases have led to a dearth of real-world computer models that are freely available and/or accessible to statisticians that hope to develop novel methods for the computer modeling field. It is for this reason that we have developed the {\bf{CompModels}} package which serves as a repository of pseudo computer models for statistical use.

The {\bf{CompModels}} package can be used to test and develop methods for computer model emulation (prediction), uncertainty quantification, and calibration, however, the main focus when developing the package was placed on building computer models for optimization. Real-world computer models are often built with the goal of understanding some physical system of interest, and with that goal usually comes the need to optimize some output of interest from the computer model. For example, in hydrology the minimization of contaminants in rivers and soils is of interest and so computer models representing pump-and-treat remediation plans are often used in order to optimize objectives, such as the associated costs of running pumps for pump-and-treat remediation, while also ensuring that contaminants do not spread \citep{pourmohamad:lee:2019}. Recalling that most computer models are computationally expensive to run, the need for efficient sequential optimization algorithms (also known as Bayesian optimization) that do not require many functional evaluations is high, which is why the focus of the test functions in the {\bf{CompModels}} package is placed on optimization. More specifically, the {\bf{CompModels}} package presents functions to optimize of the following form
\begin{align}\label{optim:problem}
\min_x & \{f(x): c(x)\leq 0,\, x\in \mathcal{X}\}
\end{align}
where $\mathcal{X}\subset \mathbb{R}^d$ is a known, bounded region such that $f:\mathcal{X}\rightarrow\mathbb{R}$ denotes a scalar-valued objective function and $c:\mathcal{X}\rightarrow\mathbb{R}^m$ denotes a vector of $m$ constraint functions. However, many of the package functions omit the constraint functions and thus the package is a mix of constrained and unconstrained optimization problems.

Some of the functions in the {\bf{CompModels}} package have known functional forms, for example the \texttt{gram()} and \texttt{mtp()} functions, however most all functions are intended to serve as black-box computer models. All of the black-box computer model functions within the package are aptly named \textsf{bbox} (short for black-box) and followed by a unique integer value to make the functions discernible. For example, \texttt{bbox1()} and \texttt{bbox2()} are two unique function calls to two different black-box computer models that can be used for constrained and unconstrained optimization, respectively. \textsf{R} is an open source programming language and so none of the computer models within the package can ever truly be a completely black-box function, however, the developers of the  {\bf{CompModels}}  package have done their best to obscure the analytical forms of the mathematical functions underlying the computer models. For example, at the first level of the code, a call to the \texttt{bbox1()} function tells the user the following:  
\begin{verbatim}
R> bbox1
function(x1,x2){

  if(!is.numeric(x1) | !is.numeric(x2) | length(x1) != 1 | length(x2) !=1){
    stop("Input is invalid.")
  }else if(x1 < -1.5  | x1 > 2.5 | x2 < -3 | x2 > 3){
    stop("Input is outside of the domain.")
  }else{
    ans <- .C("bbox1c",x1=x1,x2=x2,fx=0,c1x=0,c2x=0)
    return(list(obj = ans$fx, con = c(ans$c1x,ans$c2x)))
  }
}
\end{verbatim}
The only discernible information that the user can glean from this output is that the \texttt{bbox1()} function has an input dimension of $d=2$, where the domain $\mathcal{X} = [-1.5,2.5]\times[-3,3]$, and that there is one objective function, \texttt{fx}, to minimize, and two constraint functions, \texttt{c1x} and \texttt{c2x}, to satisfy. As we see from the \texttt{.C()} command, the actual source code for the black-box function has been written using the  \textsf{C} programming language. The \textsf{C} programs are publicly available, however,  the code within those programs has been heavily obfuscated to the best of our abilities in order to obscure the source code such that the computer models remain black-box functions. Moreover, we believe that good robust methodology developed for computer models benefits from being applied to black-box functions and so any attempt to decipher the black-box computer models is simply a disservice to the statistician developing the methodology.

\begin{figure}[!http]
\centering
\includegraphics[scale=.3]{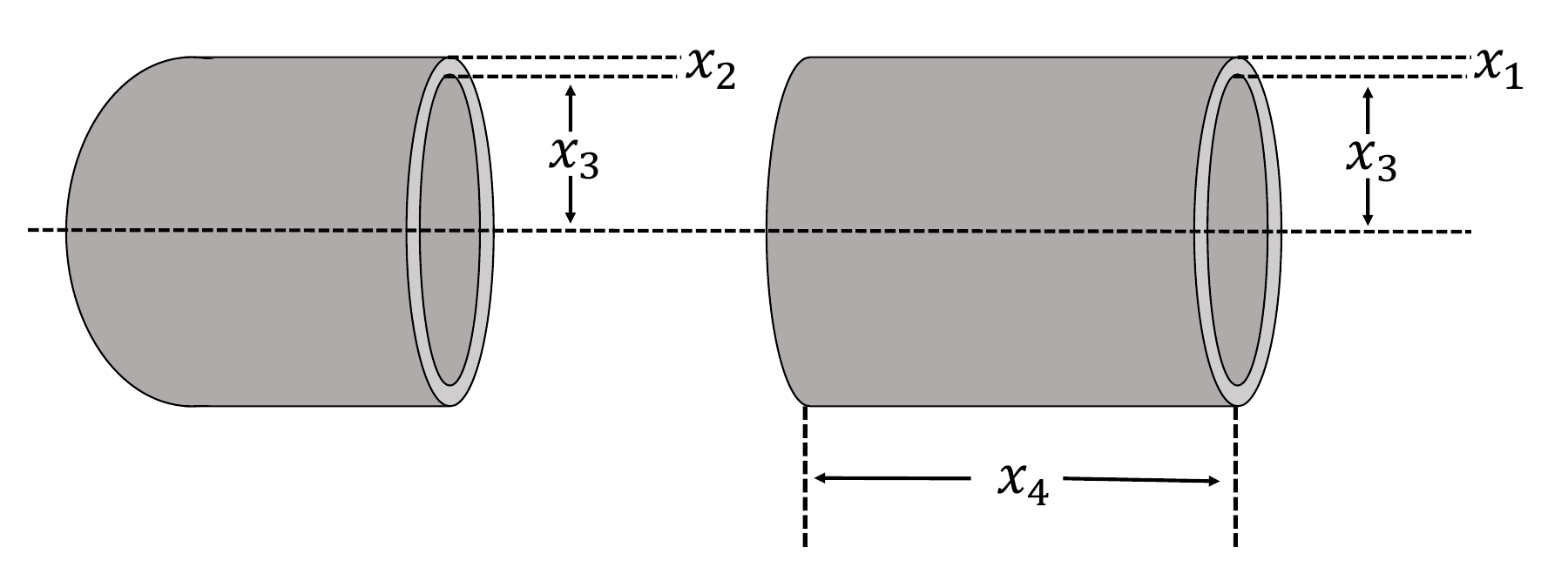}
\caption{\label{fig:pressure} The physical representation of the pressure vessel computer model.}
\end{figure}

When developing the computer models in the package we kept in mind that the best computer model examples typically have roots in real applications. When possible, we tried to develop computer models that were either based in physics, or that appeared in the literature with real use cases. For example, one computer model, \texttt{pressure()}, is based on the real-world engineering problem of minimizing the cost associated with constructing a pressure vessel (Figure \ref{fig:pressure}). Given the thickness of the shell ($x_1$), the thickness of the head ($x_2$), the inner radius ($x_3$), and the length of the cylindrical section of the vessel ($x_4$) not including the head, the cost of constructing the pressure vessel is to be minimized subject to four constraints on the cost of materials, forming, and welding.

\begin{figure}[!http]
\centering
\includegraphics[scale=.3]{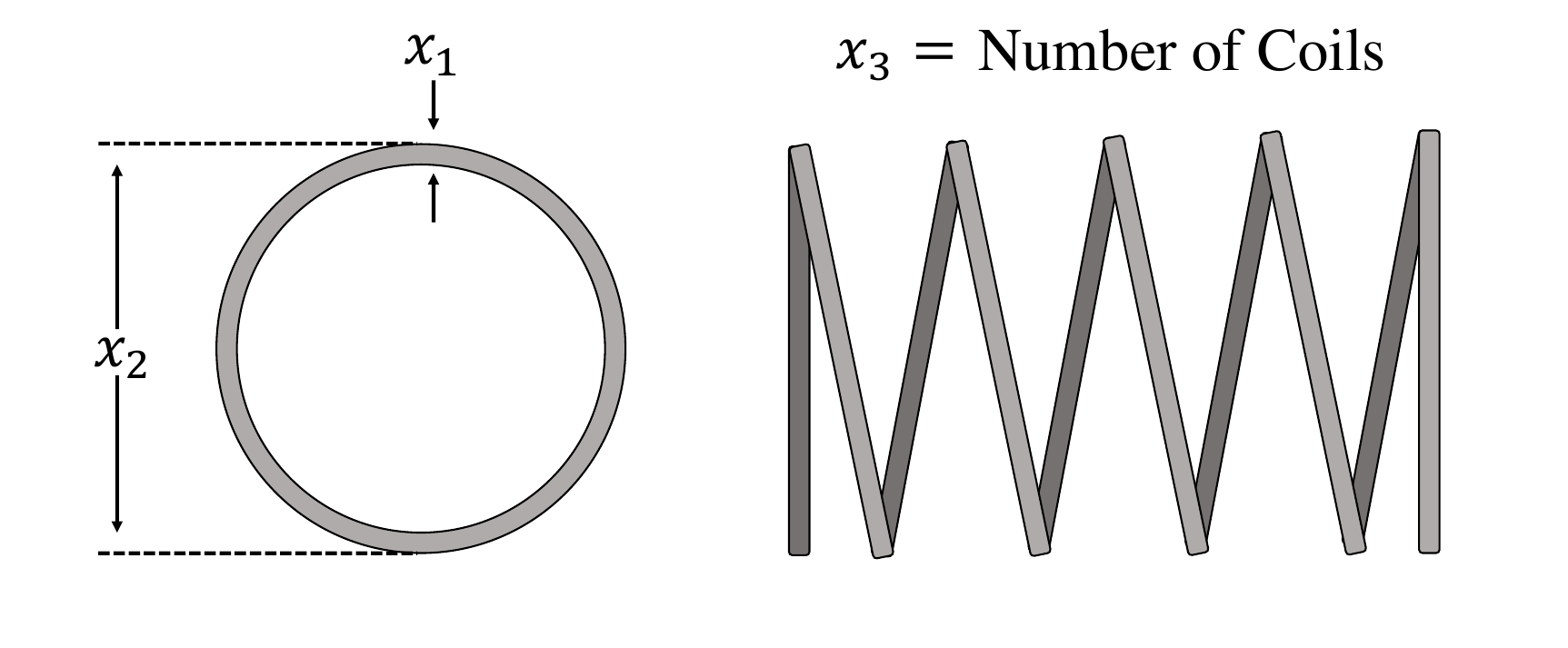}
\caption{\label{fig:tension} The physical representation of the tension spring computer model.}
\end{figure}

Likewise, when possible, we sought out real-world problems where solutions already existed that could be benchmarked to. For example, the tension spring computer model, \texttt{tension}, is designed to minimize the weight of a tension spring (Figure \ref{fig:tension}) subject to four constraints on the shear stress, surge frequency, and deflection. The three inputs to the computer model are for the wire diameter ($x_1$), mean coil diameter ($x_2$), and the number of active coils ($x_3$). 

The tension spring problem has been solved many times in the literature, and Table \ref{tab:tension} summarizes some of the best solutions.

\newpage
\begin{table}[!http]
\footnotesize
\centering
\begin{tabular}{lcccr}
\hline
& \multicolumn{3}{c}{Optimal Inputs} & \\
\cline{2-4}
Source  & $x_1$ & $x_2$ & $x_3$ & Best Solution\\
\hline
 \cite{coello:2000}& 0.051480 & 0.351661 & 11.632201 & 0.012704\\
\cite{he:2007}& 0.051728 & 0.357644 & 11.244543 & 0.012675\\
 \cite{gandomi:2013}&  0.051690 & 0.356730 & 11.288500 & 0.012670\\
\cite{mirjalili:2014}& 0.051690 & 0.356737 & 11.288850 & 0.012666\\
\cite{lee:2005}& 0.051154 & 0.349871 & 12.076432 & 0.012671\\
\cite{askarzadeh:2016}&  0.051689 & 0.356717 & 11.289012 & 0.012665\\
\cite{mirjalili:2017}& 0.051207 & 0.345215 & 12.004032 & 0.012676\\
\cite{li:2019}& 0.051618 & 0.355004 & 11.390144 & 0.012665\\
\hline
\end{tabular}
\caption{\label{tab:tension} Best solutions to the tension spring optimization problem from the literature.}
\end{table}

We stress the need for benchmarking in our examples because we believe that benchmarking also helps with allowing for good computer model methodology to be developed. Often times in the computer modeling literature you tend to see real-world optimization results that stand alone and cannot be compared against, or even replicated, because practitioners do not have access to the same computer models as others. Being able to benchmark one's results to others helps discern how well a given optimization method performs, as well as allowing for useful internal feedback when developing a method. Thus, a key reason we have developed the {\bf{CompModels}} package is so that equitable access to computer models for benchmarking exists. Similarly, a problem with real-world computer models is that they can change over time and often older versions will be phased out, unsupported, or disappear entirely. For example, the optimization results for the MODFLOW-96 computer model \citep{mcdonald:1996} from \cite{pourmohamad:lee:2016} was benchmarked to the work in \cite{lindberg:lee:2015}, however, this computer model is no longer supported by its developers and so future benchmarking may become infeasible. Thus the {\bf{CompModels}} package also stands as a repository of computer models that should be available to all users for the foreseeable future. Lastly, computer models can often times be platform and operating system specific which ultimately limits the number of potential users of the computer model. Given that \textsf{R} packages, for the most part, tend to be immune to this problem, the {\bf{CompModels}} package would be available to as wide of an audience as possible, again providing equitable access to computer models.

The remainder of the paper is organized as follows. Section \ref{sec:BO} gives a brief introduction to Bayesian optimization and expected feasible improvement so that the computer models within the {\bf{CompModels}} package can be demonstrated. Section \ref{sec:illustrations} illustrates practical applications of package use for optimization, and Section \ref{sec:disc} concludes with a discussion.

\section{Bayesian Optimization} \label{sec:BO}
Tracing its roots as far back as to \cite{mockus:1978}, Bayesian optimization (BO) is a sequential design strategy for efficiently optimizing black-box functions, in few steps, that does not require gradient information \citep{brochu:2010}. More specifically, BO seeks to solve the minimization problem 
\begin{align}\label{BO}
x^* = \argmin_{x\in\mathcal{X} }f(x).
\end{align}
The minimization problem in (\ref{BO}) is solved by iteratively developing a statistical surrogate model of the unknown objective function $f$, and at each step of this iterative process, using predictions from the statistical surrogate model to maximize an acquisition (or utility) function, $a(x)$, that measures how promising each location in the input space, $x\in\mathcal{X}$, is if it were to be the next chosen point to evaluate. As alluded to in Section \ref{sec:intro}, the GP is the typical choice of surrogate model in the computer modeling literature,  and so we adopt that stance as well in this paper. Lastly, although the general definition of BO is that of an unconstrained optimization problem, extensions to the constrained optimization case are straightforward and many \citep{lee:gramacy:linkletter:gray:2011, gramacy:gray:digabel:lee:ranjan:wells:wild:2016, letham:2019, pourmohamad:lee:2020}. Here, we merely augment the original problem statement in (\ref{BO}) to be 
\begin{align}\label{CBO}
x^* = \argmin_{x\in\mathcal{X} }f(x)\,\,\,\text{subject to}\,\ c(x)\leq0,
\end{align}
where now both $f$ and $c$ can be modeled using independent GPs, and all other steps proceed as before.

In order to solve the problems in (\ref{BO}) and (\ref{CBO}), an acquisition function must be chosen for efficiently guiding the search. Perhaps one of the most popular acquisition functions for unconstrained Bayesian optimization is that of expected improvement (EI) \citep{jones:schonlau:welch:1998}. Originally introduced in the computer modeling literature, \cite{jones:schonlau:welch:1998} defined the improvement statistic at a proposed input $x$ to be $I(x)=\max_x\{0,f_{\min}^n-Y(x)\}$ where, after $n$ runs of the computer model, $f^n_{\min}=\min\{f(x_1),...,f(x_n)\}$ is the current minimum value observed. Since the proposed input $x$ has not yet been observed, $Y(x)$ is unknown and can be regarded as a random variable. Likewise, $I(x)$ can be regarded as a random variable and so new candidate inputs, $x^*$, can be selected by maximizing the expected improvement, i.e.,
\begin{align}\label{expected:improvement}
x^*=\arg\max_{x\in\mathcal{X}}\mathbb{E}[I(x)].
\end{align}
Fortunately, if we treat $Y(x)$ as coming from a GP then, conditional on a particular parameterization of the GP, the EI acquisition function is available in closed form as
\begin{align}\label{EI}
\mathbb{E}[I(x)] = (f^n_{\min} - \mu^n(x))\Phi\left(\frac{f_{\min}^n-\mu^n(x)}{\sigma^n(x)}\right) + \sigma^n(x)\phi\left(\frac{f_{\min}^n-\mu^n(x)}{\sigma^n(x)}\right),
\end{align}
where $\mu^n(x)$ and $\sigma^n(x)$ are the mean and standard deviation of the predictive distribution of $Y(x)$, and $\Phi(\cdot)$ and $\phi(\cdot)$ are the standard normal cdf and pdf respectively.  

Extending EI to the constrained optimization case, \cite{schonlau:1998} defined expected feasible improvement (EFI) as 
\begin{align}
\text{EFI}(x) = \mathbb{E}[I(x)]\times\text{Pr}(c(x)\leq0),
\end{align}
where  $\text{Pr}(c(x)\leq 0)$ is the probability of satisfying the joint constraints. Here, $I(x)$ uses an $f^n_{\min}$ defined over the region where the constraint functions are satisfied.  Again, new candidate inputs, $x^*$, can now be selected by maximizing the expected feasible improvement, i.e.,
\begin{align}\label{expected:feasible:improvement}
x^*=\arg\max_{x\in\mathcal{X}} \mathbb{E}[I(x)]\times\text{Pr}(c(x)\leq0). 
\end{align}
Here the formula in (\ref{EI}) still holds, however, we are now weighting EI by the probability that $x$ is feasible.

\section{Illustrations} \label{sec:illustrations}
We illustrate the use and functionality of the computer models in the {\bf{CompModels}} package by solving two constrained optimization problems using the EFI method outlined in Section \ref{sec:BO}. We optimize the tension spring computer model, \texttt{tension()}, as well as the black-box 1 computer model, \texttt{bbox1()}. In both cases, we perform Monte Carlo experiments where we repeat the optimization routine a total of 30 times to judge the robustness of the solutions. We take advantage of the function \texttt{optim.efi()} in the{\bf{laGP}} package \citep{laGP} for running the EFI algorithm. A full list of the available computer models in the {\bf{CompModels}} package is given in Appendix \ref{appendix}, and are generalizable to the proceeding examples.

\subsection{Tension Spring Computer Model}
The goal of the tension spring computer model is to minimize the weight of the tension spring subject to four constraints on the shear stress, surge frequency, and deflection. Here, the inputs to the tension spring computer model are the wire diameter ($x_1$), mean coil diameter ($x_2$), and the number of active coils ($x_3$), where $x_1\in[0.05,2]$, $x_2\in[0.25,1.3]$, and $x_3\in[2,15]$. To evaluate the computer model at a given input, a user needs to supply an input within the given domain, i.e., 
\begin{verbatim}
R> tension(x1 = 1, x2 = 1, x3 = 3)
$obj
[1] 5

$con
[1]   0.9999582 -45.8166667  -0.9995655   0.3333333
\end{verbatim}
All of the computer model functions in the package will return a list where the first element in the list is the value of the objective function, and (in the case of constrained optimization) the second element contains the values of the constraint functions. Here we see that for a wire diameter of \texttt{x1 = 1} , mean coil diameter of \texttt{x2 = 1} , and \texttt{x3 = 3}  active coils, that the weight of the tension spring is five, however, the first and last constraint has not been satisfied since those values of \texttt{\$con} are non-negative. Thus, the input is not a feasible solution to the problem. The input of $x=(1,1,3)$ was merely a guess for illustrative purposes. A more reasonable approach to minimizing the tension spring computer model would be to employ the EFI method in Section \ref{sec:BO}. In order to do so, we make use of the function \texttt{optim.efi()} in the {\bf{laGP}} package. To be able to use the \texttt{optim.efi()} function, we need to first build a wrapper function (which we call \texttt{bbox}) for our \texttt{tension()} function that conforms to the specifications of the \texttt{optim.efi()} function. 
\begin{verbatim}
R> bbox <- function(X){
+  output = tension(X[1], X[2], X[3])
+  return(list(obj = output$obj, c = output$con))
}
\end{verbatim}
Next we need to create a matrix that encodes the domain of the computer model inputs. 
\begin{verbatim}
R> B <- matrix(c(.05, .25, 2, 2, 1.3, 15), nrow=3)
\end{verbatim}
We can implement the EFI algorithm by passing our wrapper function and domain variable as arguments to the \texttt{optim.efi()} function, and then by checking the regions where the solution satisfies the constraints. 
\begin{verbatim}
R> ans <- optim.efi(bbox, B, fhat = TRUE, start = 10, end = 300)
R> constraint <- ifelse(apply(ans$C, 1, max) > 0, "Not Met", "Met")
\end{verbatim}
Here we see that the \texttt{optim.efi()} function started with a random input of 10 data points and sequentially chose 290 more inputs for a total of 300 evaluations. The output of \texttt{optim.efi()} is a large list storing all steps of the EFI algorithm. We create the \texttt{constraint} variable in order to be able to find where the minimum feasible value exists. 
\begin{verbatim}
R> min(ans$obj[constraint == "Met"])
[1] 0.0112376

R> ans$X[ans$obj == min(ans$obj[constraint == "Met"])]
[1] 0.05345441 0.45253754 6.69064005
\end{verbatim}
Here we see that the best feasible value found by the EFI algorithm is at a weight of 0.0112376 which occurs at an input of $x=(0.05345441,0.45253754,6.69064005)$. Interestingly, this minimum value found of 0.0112376 is much smaller than all of the best minimums found in our review of the literature (Table \ref{tab:tension}). To evaluate the robustness of the EFI algorithm for the tension spring computer model, we conduct a Monte Carlo experiment where we repeat the optimization routine 30 times based on different starting input data sets of size 10. 
\begin{verbatim}
R> S <- 30 
R> results <- rep(NA, S)
R> for(i in 1:S){
+  ans <- optim.efi(bbox, B, fhat = TRUE, start = 10, end = 300)
+  constraint <- ifelse(apply(ans$C, 1, max) > 0, "Not Met", "Met")
+  results[i] <- min(ans$obj[constraint == "Met"])
+}

R> summary(results)
   Min. 1st Qu.  Median    Mean 3rd Qu.    Max. 
0.01081 0.01255 0.01302 0.01325 0.01386 0.01859 
\end{verbatim}
From the summary of the results, we see that over the 30 Monte Carlo experiments that the EFI algorithm was not able to reliably find as good of a solution over the 300 computer model evaluations. The mean value over the 30 runs was 0.01325 which was much higher than the best solutions presented in Table \ref{tab:tension}. However, we do see from the summary that the EFI algorithm was able to find at least one more better solution, as compared to the literature, at a spring weight of 0.01081. 

\subsection{Black-box Computer Model}
\begin{figure}[!http]
\centering
\includegraphics[scale=.5]{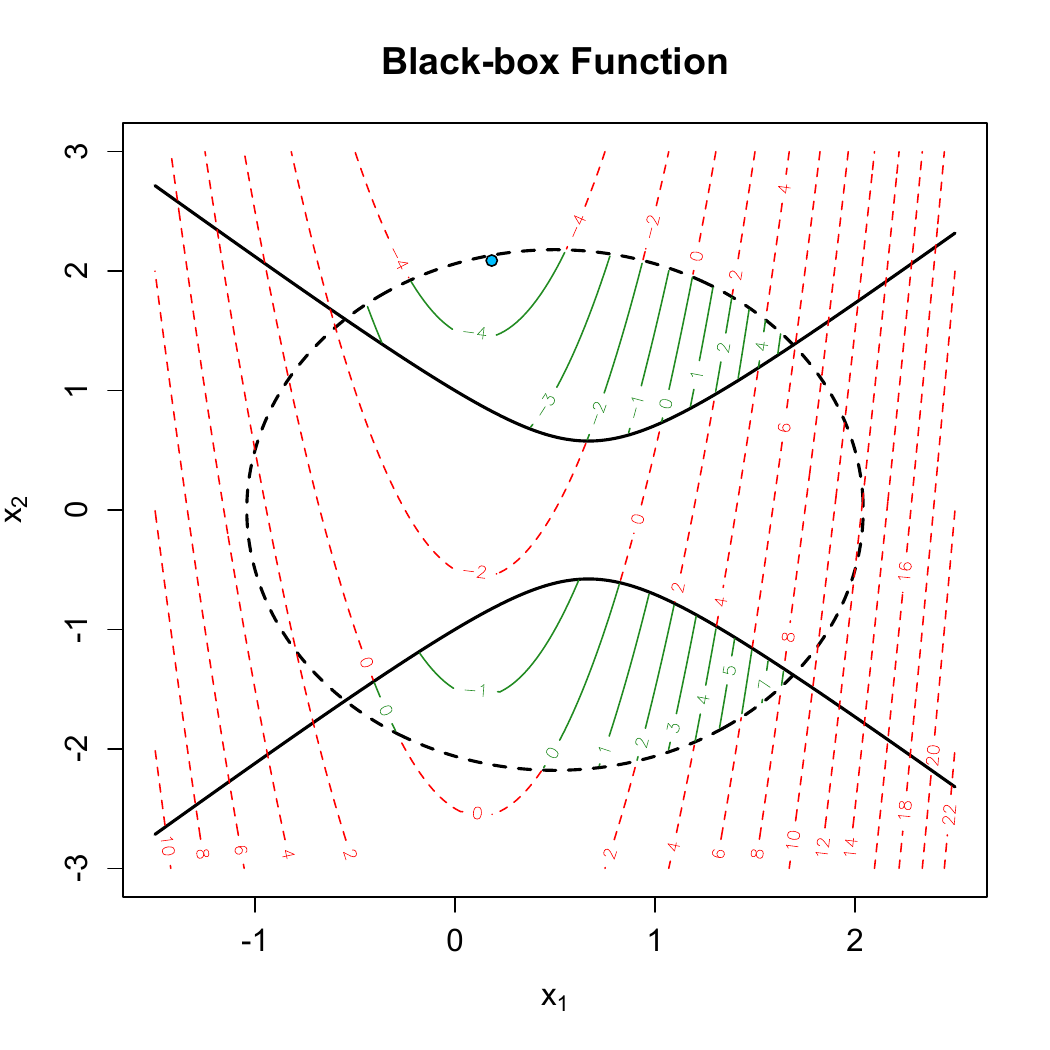}
\caption{\label{fig:blackbox} The objective function colored by the two constraints. The solid black line denotes one constraint function, while the dashed black line denotes the other constraint function. Contours that are red are areas where the constraints are not satisfied, while green contours indicate areas where the constraints are satisfied. The blue point represents the best feasible solution found by EFI.}
\end{figure}

Recalling Section \ref{sec:intro}, the \texttt{bbox1()} computer model has an input dimension of $d=2$, where the domain $\mathcal{X} = [-1.5,2.5]\times[-3,3]$, and that there is one objective function, \texttt{fx}, to minimize, and two constraint functions, \texttt{c1x} and \texttt{c2x}, to satisfy. We can once again use the  \texttt{optim.efi()} function to perform the EFI algorithm by creating an appropriate wrapper function and domain variable.
\begin{verbatim}
R> bbox <- function(X){
+  output = bbox1(X[1], X[2])
+  return(list(obj = output$obj, c = output$con))
+}

R> B <- matrix(c(-1.5, -3, 2.5, 3), nrow = 2)
\end{verbatim}
We initialize the \texttt{optim.efi()} function with an input data set of 10 points and continue to sequentially evaluate the \texttt{bbox1()} function for a total of 100 input points.
\begin{verbatim}
R> ans <- optim.efi(bbox, B, fhat = TRUE, start = 10, end = 100)
R> constraint <- ifelse(apply(ans$C, 1, max) > 0, "Not Met", "Met")
\end{verbatim}
 Checking the EFI  algorithm results in the areas where the constraint functions were satisfied we obtain a best feasible minimum objective function value of -4.61008 which occurs at $x=(0.204649,2.072964)$.
\begin{verbatim}
R> min(ans$obj[constraint == "Met"])
[1] -4.610088

R> xbest <- ans$X[ans$obj == min(ans$obj[constraint == "Met"])]
R> xbest
[1] 0.204649 2.072964
\end{verbatim}
Now, since the \texttt{bbox1()} function is a black-box computer model, we do not have any analytical way of checking whether or not our solution to the optimization problem is a good one. However, the functions in the {\bf{CompModels}} package were not developed with the intent of forcing them to be computationally  expensive if they need not be. Thus, with an input dimension of $d=2$, it is very easy to evaluate the  \texttt{bbox1()} function on a very dense grid to understand what the potential surface of the objective and constraint functions look like. Doing so does not guarantee us analytically that our solution is a good one, however, we will be able to tell visually whether or not our solution is a good one. Plotting the objective and constraint surfaces we obtain the following (Figure \ref{fig:blackbox}).
\begin{verbatim}
R> n <- 200
R> x1 <- seq(-1.5, 2.5, len = n)
R> x2 <- seq(-3, 3, len = n)

R> x <- expand.grid(x1, x2)
R> obj <- rep(NA, nrow(x))
R> con <- matrix(NA, nrow = nrow(x), ncol = 2)

R> for(i in 1:nrow(x)){
+  temp <- bbox1(x[i,1], x[i,2])
+  obj[i] <- temp$obj
+  con[i,] <- temp$con
+}

R> y <- obj
R> y[con[,1] > 0 | con[,2] > 0] <- NA

R> z <- obj
R> z[!(con[,1] > 0 | con[,2] > 0)] <- NA

R> par(ps=15)
R> plot(0, 0, type = "n", xlim = c(-1.5, 2.5), ylim = c(-3, 3), 
+     xlab = expression(x[1]), ylab = expression(x[2]), main = "Black-box Function")
R> c1 <- matrix(con[,1], ncol = n)
R> contour(x1, x2, c1, nlevels = 1, levels = 0, drawlabels = FALSE, add = TRUE, 
+     lwd = 2)
R> c2 <- matrix(con[,2], ncol = n)
R> contour(x1, x2, c2, nlevels = 1, levels = 0, drawlabels = FALSE, add = TRUE, 
+    lwd = 2, lty = 2)
R> contour(x1, x2, matrix(y, ncol = n), nlevels = 10, add = TRUE, col = "forestgreen")
R> contour(x1, x2, matrix(z, ncol = n), nlevels = 20, add = TRUE, col = 2, lty = 2)
R> points(xbest[1], xbest[2], pch = 21, bg = "deepskyblue")
\end{verbatim}

By plotting the objective function surface, along with the constraint functions, see that the space where the constraints are satisfied are two disconnected regions where the feasible region with $x_1>0$ has much lower objective function values than the feasible region where $x1<0$. We plotted our best minimum objective value found, by EFI, as a blue circle in (Figure \ref{fig:blackbox}). Visually, our best minimum objective value found appears to be around the global minimum value based on the calculated contour lines of the plot. Although this visual inspection suggests that our EFI algorithm has correctly identified the global solution to the optimization problem, confirmation of our solution could come from others using the {\bf{CompModels}} package in order to benchmark the solution. Lastly, we check the robustness of the solution found by EFI algorithm by conducting a Monte Carlo experiment where we repeat the optimization routine for a total of 30 times.
\begin{verbatim}
R> summary(results)
   Min. 1st Qu.  Median    Mean 3rd Qu.    Max. 
 -4.681  -4.661  -4.645  -4.645  -4.632  -4.602
\end{verbatim}
From the summary of the results, we see that the variation in the results show up in the hundredth decimal point, and beyond, which we regard as representing a very robust solution. 

\section{Discussion} \label{sec:disc}
The primary goal of the package is to provide users a source of computer model test functions that are reproducible, shareable, and that can ultimately be used for benchmarking of Bayesian optimization methods. The package will greatly benefit those who do not have access, or connections, to real-world computer models. In time, it is our hope that the package will come to be viewed as a suite of real computer models rather than solely as pseudo ones. Likewise, the {\bf{CompModels}} package is a not a static package in that we envision it to be a living repository, and so more computer model functions will be expected to be added over time. The success of any \textsf{R} package ultimately comes from the feedback received from its users. We greatly encourage all interested users of the package to please contact the developers in order to provide any insights or examples for new computer models to be added. 

\bibliographystyle{apalike}
\bibliography{refs}

\appendix
\section{Current Computer Models}\label{appendix}
Table \ref{tab:append} provides a summary of the current computer models that are available in the {\bf{CompModels}} package. The package is a mix of real-world physics problems, known mathematical functions, and black-box functions, as well as a mix of constrained or unconstrained optimization problems.

\begin{table}[!http]
\centering
\begin{tabular}{lccc}
\hline
Function  & Input Dimension & Optimization Type & No. of Constraints \\
\hline
\texttt{bbox1()} &  2 & Constrained & 2 \\
\texttt{bbox2()} & 2 & Unconstrained & -- \\
\texttt{bbox3()} & 2 & Unconstrained & -- \\
\texttt{bbox4()} & 2 & Constrained & 1 \\
\texttt{bbox5()} & 3 & Unconstrained & -- \\
\texttt{bbox6()} & 1 & Constrained & 2 \\
\texttt{bbox7()} & 8 & Constrained & 2 \\
\texttt{gram()} & 2 & Constrained & 2 \\
\texttt{mtp()} & 2 & Constrained & 2 \\
\texttt{pressure()} & 4 & Constrained & 4 \\
\texttt{sprinkler()} & 8 & Unconstrained & -- \\
\texttt{tension()} & 3 & Constrained & 4 \\
\hline
\end{tabular}
\caption{\label{tab:append} Current computer models that are implemented in the {\bf{CompModels}} package.}
\end{table}

\end{document}